# Low-frequency internal friction (LFIF) as express-method for identification of cryocrystals in pores of the solids

A.I. Erenburg[1], A.Yu. Zakharov[2], A.V. Leont'eva[3], A.Yu. Prokhorov[3]

[1] Ben-Gurion University of Negev, P.O.B. 653, Beer-Sheva 84105, Israel
  E-mail: erenburga@gmail.com, erenbura@bgu.ac.il
[2] Novgorod State University, B. S.-Peterburgskaya str. 41, Velikiy Novgorod 173003, Russia
  E-mail: anatoly.zakharov@novsu.ru
[3] Donetsk Institute for Physics & Engineering, NAS of Ukraine, Prospect Nauki 46, Kiev, 03680, Ukraine; E-mail: vesta-news@yandex.ru, tonya.leont@gmail.com

## Abstract.

We show that studying of low-frequency internal friction (LFIF) of solid samples at low temperatures allows determining the presence of various gases absorbed, for some reasons, in pores and caverns of the solids. The gases come over to a solid state (cryocrystals) and exist in the pores under corresponding thermodynamic conditions giving an additional contribution to the LFIF spectra. The spectra reflect the special points of the gases (temperatures of melting or phase transitions). This information gives a real opportunity for identification of gas in the matrix, i.e. the studied solids. This may be of great importance for investigations of cosmic or geological samples, for instance, asteroids, meteorites, rock formations, etc. The LFIF method allows identification of gas media surrounding the studied sample.

## Introduction.

Cryocrystals (or solidified gases) form a relatively small group of materials which are gaseous at room temperatures and solids at low ones. They have triple points at low temperatures because of low weight and small size of their molecules and also weakness of binding forces. This group of solids includes atomic cryocrystals (He, Ne, Ar, Kr, Xe), simplest molecular crystals (hydrogen, nitrogen, CO, oxygen, and also few crystals from larger molecules ($CO_2$, $N_2O$, $CH_4$, $NH_3$ etc.) [1]. Being chemically neutral, the gases fill the pores and becomes cryocrystals under cooling, influencing the properties of solid matrices.

Method of internal friction used in the paper permits to detect a presence of gases in pores of the matrix as the cryocrystals give an additional contribution in background spectrum of crystal matrix in the form of additional peaks. Temperatures of triple points are the control points for most of the cryocrystals. For molecular cryocrystals, the control points include the temperatures of phase transitions as well.

## I. Problems of oxide ceramics and determination of gaseous oxygen existence in its pores

Oxide ceramics are usually fabricated by sintering and have numerous internal pores. These pores can absorb various gases influencing the mechanical, thermodynamic or other properties of the materials.

Investigations of gas media influence on the properties of the solids represent a very extensive area. It includes the studies of influence of gas solved in the solid, gas in pores and gas condensed on the surface, on various characteristics of the ceramics. A problem of influence of oxygen media on various properties of the ceramics can be analyzed separately. For instance, it was shown in [2] that oxygenation of HIP-produced silicon nitride during 300 h at T=1400 C leads to more than 50% increase of its strength due to the formation on its surface of a film enriched by oxygen.

Recently, a number of papers were published on problems of oxide ceramics $La_{1-x}A_xMnO_3$ (A=Ca, Sr, Ba) or manganites [3,4]. These have attracted a great interest due to existence of colossal magnetic resistance effect. At last time, such ceramics are often used for producing of thin film ferroelectric capacitors where it they play a role of magnetic component [5]. Many of oxide ferroelectric compounds, i.e. BiFeO3, BiMnO3, along with HTSC ceramics (LSMO, YBCO) are nonstoichiometric materials in relation to $O_2$. They are often additionally oxygenated, for instance during the thermal treatment to improve its ferroelectric or superconducting characteristics [6]. So, thin films of well-known ferroelectrics BiFeO3 have a problem of existence of significant leak currents, caused by, among others, number of oxygen vacancies forming during the film growth. Annealing in oxygen media is one of the ways for solving this problem [7,8]. However, such thermal treatment often changes the characteristics of the compound.

Therefore, the determining of oxygen content in ceramics is a crucial task. In this connection, our data on the influence of oxygen in pores on the low temperature internal friction spectra are of great scientific and applied interest because they have been obtained by a non-destructive method.

Let us consider this problem by the example of studies of HTSC ceramic samples (YBaCuO and LaSrCuO [9]).

## Experimental technique and results

It is necessary to note that available internal friction spectra can be separated on two frequency regions: low frequency ones (below 500 Hz, LFIF method) and high-frequency spectra (frequency higher than $10^3$ Hz – HFIF method). In this paper, we shall consider these methods with the point of view of its sensitivity to the presence of gases in pores of the studied solids.

The cryostat for LFIF investigation of solidified gases (cryocrystals) at temperatures from liquid helium up to room temperature is shown in Fig. 1. Details of the experimental design were published earlier [10, 11].

It is important to note that a sample of the cryocrystal grown in a special chamber was scaled off from the walls of the chamber by vacuum pumping, at this the sample bottom was remained fast connected with the chamber. Torsion loads were applied to the free part of the sample.

The LFIF was measured by method of inverse torsion pendulum under deformation amplitude $10^{-4}$–$10^{-5}$ using an original setup for study of nonelastic properties of solidified gases [11-14]. An experimental error of the LFIF measurements does not exceed 5%.

In case of HTSC ceramic samples, the chamber (15) of the cryostat in Fig.1 was replaced by two special gripes (Fig.2) for fixing the studied samples which were specially fabricated according to the conditions [11]. The ceramic samples had a form of "wire" with diameter less than 1 mm.

Like as for cryocrystals, the method of inversed torsion pendulum was used for this case as well. The magnitude of deformation was $10^{-4} – 10^{-5}$. Experimental frequency of the torsion oscillations was about 20 Hz. An experimental error of the LFIF measuring was the same as for cryocrystals and does not exceed 5%.

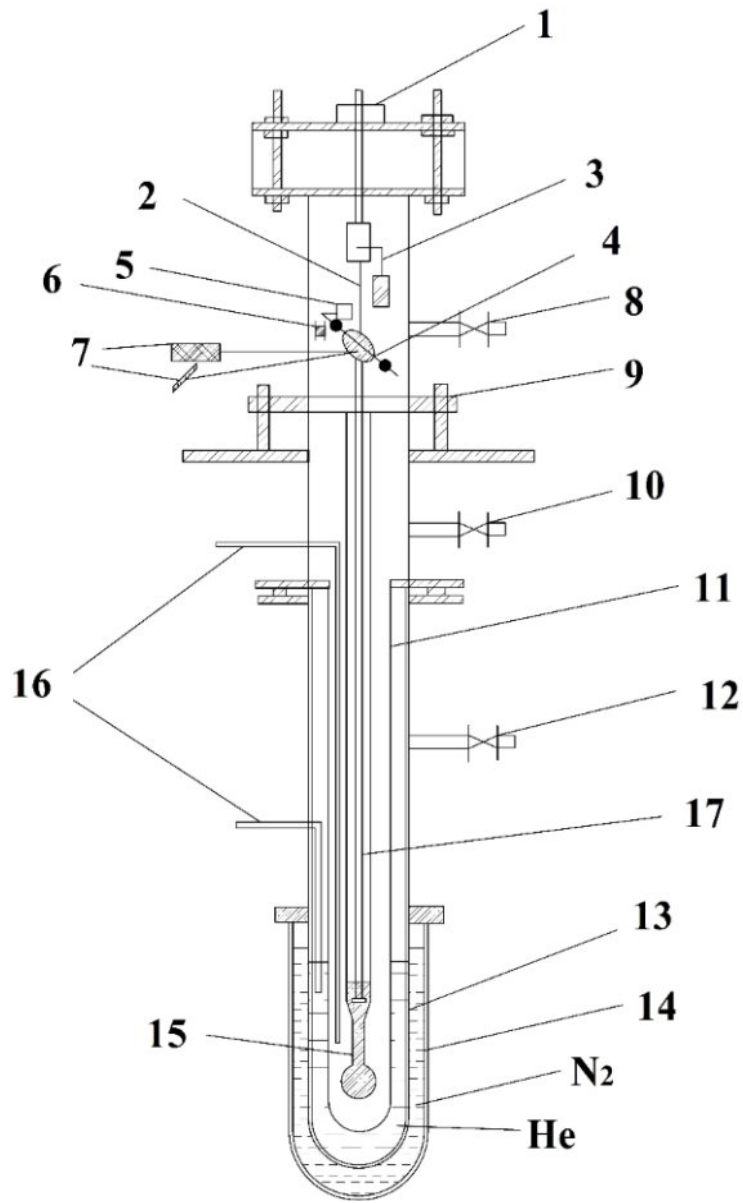

Fig. 1. Schematic of the cryostat for low frequency internal friction measurements in cryocrystals at helium temperatures:
suspension regulator for the pendulum (1); suspension wire (2); counterweight (3); internal yoke (4); inductive sensor (5); electromagnet (6); optical system (7); ports for vacuum pumping, admission of gas for study, and helium (8, 10, 12); cryostat mount (9); thermal shield (11); helium Dewar; (13); nitrogen Dewar (14); sample container (15); helium circulation tubes (16); extension of the internal system of the pendulum (17).

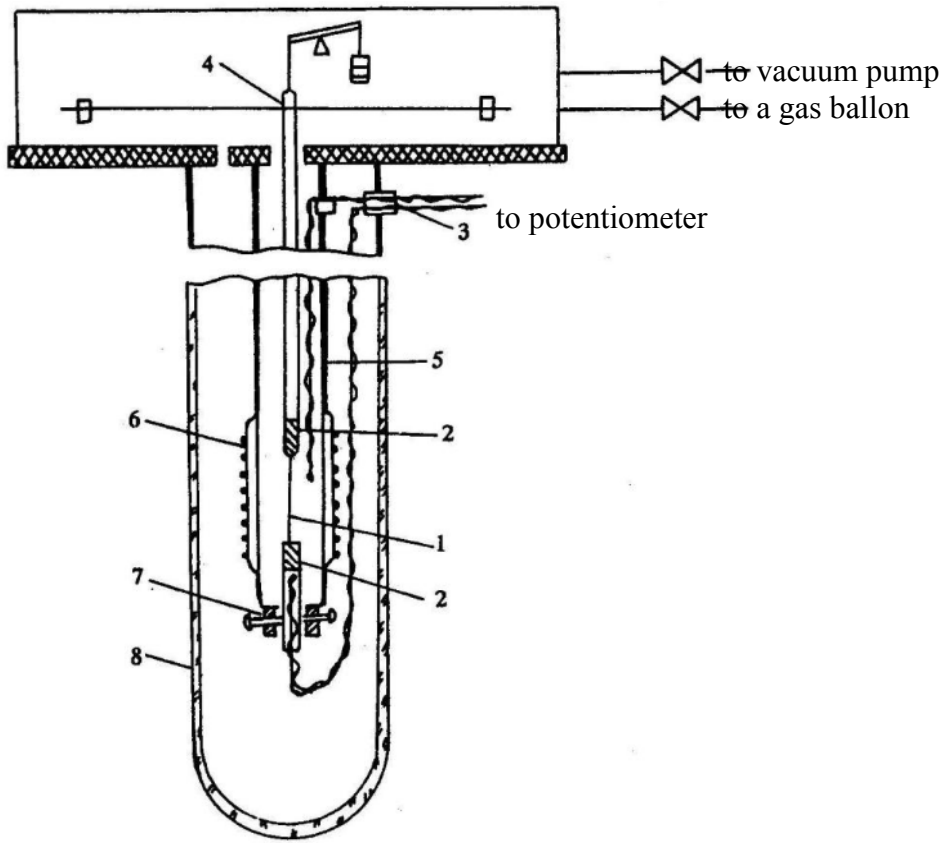

Fig.2. The operational unit of the setup for low-frequency internal friction study of ceramic samples.
(1) – sample; (2) – top and bottom extension holders of the sample; (3) – thermocouples;
(4) – torsion pendulum; (5-7) – holding elements of bottom sample adapter;
(8) – vacuum chamber for LFIF measurements in different gaseous media. Heater is placed on the element (6).

The temperature dependence of LFIF for a ceramic $YBa_2Cu_3O_{7-x}$ sample is presented in Fig.3 [16].

Fig.4 shows the results of [17, 18] where the temperature dependencies of $Q^{-1}$ have been measured by a method of composite piezoelectric vibrator. A computerized set-up was used for measuring of damping ratio $\delta$, vibration magnitude, resonance frequency and electrical resistance of a sample. Magnitude of deformation was $10^{-6}$. The experimental error did not exceed 5%.

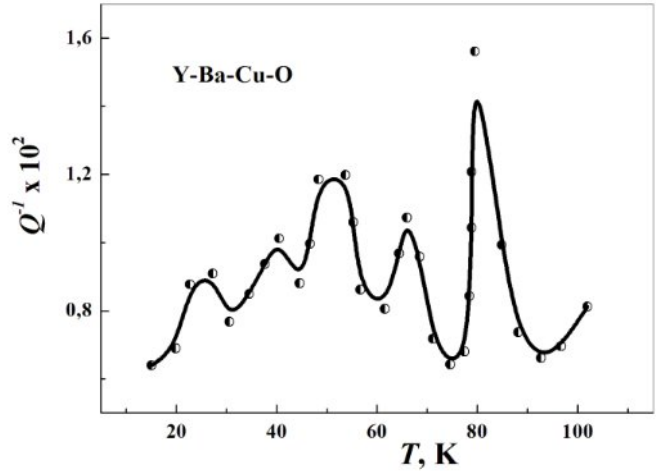

Fig.3. Temperature dependence of LFIF for a ceramic YBa$_2$Cu$_3$O$_{7-x}$ sample measured at 20 Hz [16].

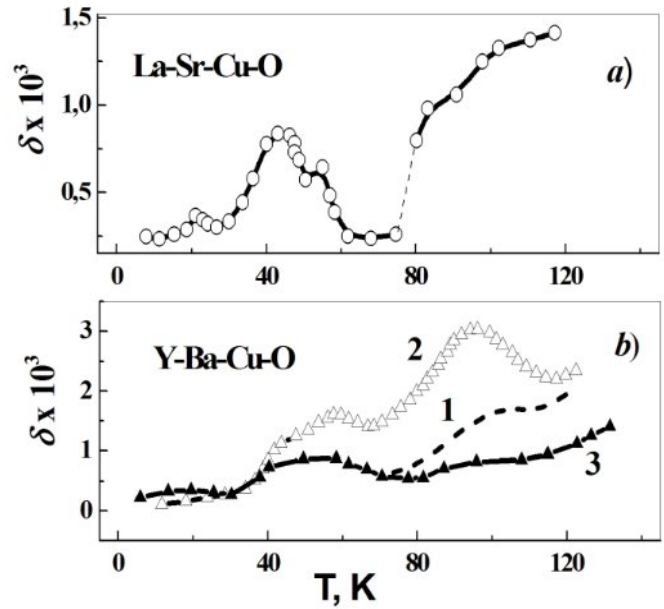

1.

Fig.4. Temperature dependencies of HFIF for lanthanum-strontium La$_{1.8}$Sr$_{0.2}$CuO$_4$ ceramic samples (a) and yttrium-barium YBa$_2$Cu$_3$O$_{7-x}$ ones (b), measured at 60 kHz [17] (a) and 100 kHz [18] (b). Lines 1-3 in Fig.4 (b) corresponds to three kinds of the samples, prepared by different methods (1, 2 – "dry" method, 3 – "wet" method).

## Estimation of sensitivity of the LFIF method

As it is seen from Figures 3-4, the LFIF peaks are observed for all the curves near the temperatures 24, 44 and 54 K. It is important to note that these peaks are more visible for LFIF spectra in comparison with the HFIF ones.

To explain this phenomenon, let us consider a simple model of damped oscillations with one degree of freedom which is described by the equation:

$$\frac{d^2x(t)}{dt^2} + \alpha\frac{dx(t)}{dt} + \omega_0^2 x(t) = 0, \qquad (1)$$

where α is the friction coefficient, $\omega_0$ is the oscillation frequency.
General solution of this equation is

$$x(t) = C\exp(-\alpha t)\cos\left(\sqrt{\omega_0^2 - \alpha^2}\, t + \varphi_0\right) \qquad (2)$$

The function $C\exp(-\alpha t)$ describes a dependency of oscillation magnitude on time, so a ratio of *n*-th magnitude $A_n$ to *(n+1)*-th one $A_{n+1}$ determines an inverse Q factor:

$$Q^{-1} = (1/\pi)\ln\left(\frac{A_n}{A_{n+1}}\right) = \frac{2\alpha}{\sqrt{\omega_0^2 - \alpha^2}}. \qquad (3)$$

As the oscillation frequency $\omega_0$ is assigned from "externality" and damping coefficient $\alpha$ is determined by the intrinsic (internal?) properties of a sample material, then the inverse Q-factor decreases with the frequency rise. Therefore, the LFIF is more sensitive to the studied material characteristics in comparison with other methods using higher frequencies, for instance, HFIF or ultra sound methods.

This is clearly demonstrated by an analysis of experimental data for solid methane in temperature interval $0.5\cdot T_{tr} - T_{tr}$ ($T_{tr}$ is the triple point temperature) under equilibrium vapor pressure [19-22].

Fig. 5 shows the temperature dependencies of LFIF spectra [14] (a) and sound velocities [23] (b) for solid methane.

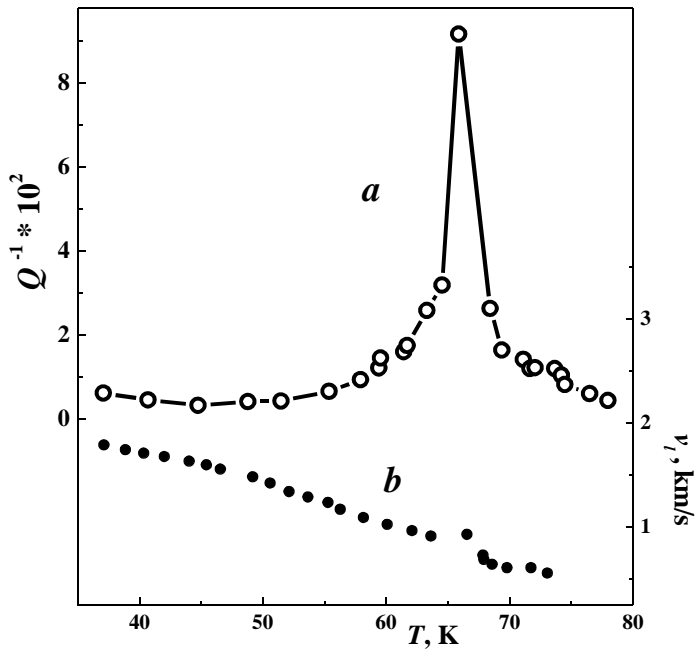

Fig.5. Temperature dependencies of LFIF $Q^{-1}$ [14] (a) and longitudinal sound velocity $v_l$ [23] (b) for solid methane.

As it is seen from Fig.5, an anomaly is observed for both curves in temperature range 60-70 K. The analyses [19-22] show that this anomaly is related to a transition of collective rotational degrees of freedom of $CH_4$ molecules from quantum character to the classical one (quantum-classical transition in a system of rotators).

Lower sensitivity of the method based on the sound velocity measurements in comparison with the LFIF method is due to that the sound velocity measurements are conducted at higher frequencies (Mega Hz range) than the measurements in the HFIF method.

## Phase transitions in solid oxygen and discussion

Since, as it was mentioned above (Figs 3-4), the sensitivity of HFIF method is significantly lower in relation to the LFIF one, it is important to note that the temperatures of $Q^{-1}(T)$ peaks (or $\delta(T)$, $Q^{-1} \sim \pi \cdot \delta$) are the same for different frequencies. This indicates that these anomalies are caused just by the phase transitions in solidified gases which fill the pores of the studied ceramic samples. A large peak near T=90 K is, probably, due to the superconducting transition in ceramics.

General investigated for HTSC ceramics, as well as other oxide ceramics is only one element - oxygen $O_2$, which is certainly present in the pores of the ceramic. Under the temperature lowering, solid oxygen in pores undergoes a few phase transitions. Temperatures of phase transitions in solid $O_2$ are near 24 and 44 K. Temperature 54 K corresponds the triple point of oxygen.

Let us consider the experimental data on phase transitions in crystalline $O_2$. According to the studies of thermal capacity [24, 25], under equilibrium vapor pressure two phase transitions are observed in solid $O_2$ at temperatures $T_{\alpha\beta}$ = 23,88 K and $T_{\beta\gamma}$ = 43,78 K. Temperature $T_{tr}$ = 54,36 K is a triple point of solid $O_2$. The structure, thermal expansion and the features of phase transitions in solid oxygen had been in detail studied in X-ray experiments [26, 27]. According to these data, a low temperature monoclinic $\alpha$ phase (space group **C2/m**) is characterized by the far orientation ordering and quasi-two-dimensional antiferromagnetic ordering of $O_2$ molecules in crystal lattice. After the transition to the mean $\beta$ phase (space group **R-3m**), only far orientation ordering of molecules in the crystal lattice is preserved. High-temperature $\gamma$ phase (space group **Pm3n**) is characterized by the total change of coordination and orientation structure of a crystal and forming of eight-molecular cubic cell. In this cell, two molecules are in fully disorientation state and other six molecules make a two-dimensional rotation around an axis which is perpendicular to the axis of the molecules. Total orientation disordering of oxygen molecules occurs in liquid and gaseous phases only.

Comparison of the **IF** temperature dependence with the data on X-ray diffraction and other thermodynamic and elastic properties indicates that the internal friction anomalies are related with the phase transitions in solid oxygen [14].

The origin of arising and storing of rather considerable amount of free oxygen in the ceramic samples (for instance, in YBa$_2$Cu$_3$O$_x$ ceramics) has been analyzed by B. Ya. Sukharevsky et.al. [15, 28, 29].

The porous samples of HTSC ceramics have been prepared from the powders of rare-earth elements using the well-known method [30]. Note that volume of pores in such polycrystalline sample may reach 10 %.

As it follows from [15, 23-24], apart from the generally known fact of oxygen losses during warmup in an air environment above 450 C, which terminates by the transition to the tetragonal phase with *q* ~ 6 (*q*=7-*x*), an oxygen release was found in the range 200 to 300 C. This process proceeds rather rapidly in ceramic (porous) samples with in this temperature range the oxygen index decreases from *q*=7 to *q*= 6.75 in a few hours which indicates a weak oxygen-lattice binding and a high mobility of the former.

As follows from Mössbauer studies [24], at *q* ≅ 7 the oxygen in the basal plane is represented in equal amounts by **O$^{2-}$** and **O$^-$** ions. This circumstance permitted the authors of Ref. 23 to propose the following oxygen release (absorption) channels:

1. Through the reaction

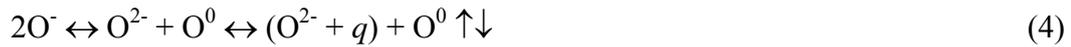

$$2O^- \leftrightarrow O^{2-} + O^0 \leftrightarrow (O^{2-} + q) + O^0 \uparrow\downarrow \qquad (4)$$

oxygen removal from the basal plane (the plane of chains) is ensured by the mobile neutral oxygen weakly bound by van der Waals forces. The "store" of O$^-$ ions is plentiful enough to let the oxygen index vary within the interval $6.75 \leq q \leq 7$.

Since the charge carrier concentration $n_0$ should follow the requirement of electrical neutrality of the crystal

$$\Sigma Z_i + n_0 = 0, \qquad (5)$$

then, with oxygen content varying via channel (1), **n$_0$** = const and, consequently, **T$_0$** = const. In Eq.(5), $n_0$ is the number of carriers per unit formula of **YBa2Cu3O$_{7-x}$**; $\Sigma Z_i$ is a sum of the ions in the unit formula of **YBa2Cu3O$_{7-x}$**. This channel is operative with q > 6.75 if the thermal treatment is "mild": heating to temperature within the 200 to 300 C range in air environment.

2. For q < 6.75, the oxygen is lost through neutralization of O$^{2-}$ ions. This is possible either because of a change of the copper ion charge

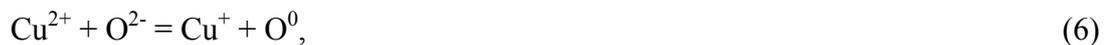

$$Cu^{2+} + O^{2-} = Cu^+ + O^0, \qquad (6)$$

or because of the decreasing of charge carrier (hole) concentration in accord with Eq. (2). This channel can be realized with a more "severe" thermal treatment: heating up to T > 440 C in vacuum or in an inert gas. The carrier concentration and the critical temperature tend to decrease here.

The amount, $N_{O_2}$ of oxygen released here per unit volume of 1 cm$^3$ of the sample can be readily estimated:

$$N_{O_2} = \Delta q/2V_z \sim 10^{21} \text{ molecules} \approx 5 \cdot 10^{-3} \text{ mole}, \qquad (7)$$

$V_z$ is the elementary cell volume; $\Delta q$ =0.25. Under normal conditions it occupies a volume $V \sim 10^2$ cm$^3$. taking the porosity of the ceramic sample to be approximately for the oxygen pressure in closed pores 10% we obtain $\sim 10^3$ atm.

Since a relation of densities for gaseous oxygen under normal conditions and condensed one below triple point is $\sim 10^3$ in order of magnitude then the vapor pressure of condensed in closed pores liquid or solid oxygen should be lower than atmospheric pressure.

Therefore, decrease of oxygen index can provide more than enough gaseous oxygen which accumulates in a ceramic "matrix" and can exist according to its own thermodynamic laws.

Procedure of preparing of "good" (with high critical temperatures and large amount of superconducting phase) samples is such that the first channel is predominantly realized so that during the slow sample cooling, i.e., during a prolonged exposure to temperatures 200°C ≤ T ≤ 300°C, the oxygen index decreases from $x \approx 7$ to x ≈ 6.75. According to our estimations, this is enough for the oxygen influence on IT spectra.

Considerable evidence for the conception of filling of internal pores in ceramic samples by oxygen is given by the experimental results of IF in single crystal YBCO samples [31]. Fig.6 shows a spectrum of ultrasound damping in a single crystal YBa$_2$Cu$_3$O$_{7-x}$ sample. As it can be clearly seen, the peaks at all characteristic points of crystal oxygen are absent (apparently due to absence of porosity in single crystal).

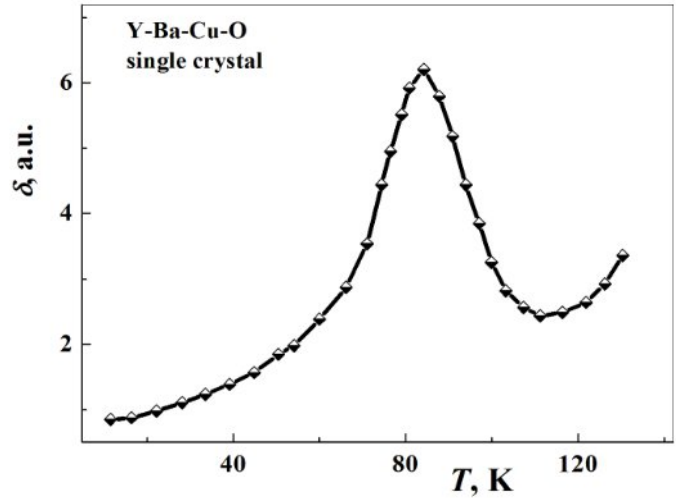

Fig.6. Temperature dependence of damping decrement δ(T) of ultrasound in a single crystal **YBa$_2$Cu$_3$O$_{7-x}$** sample according to [31].

Experimentally obtained IT spectra for pure oxygen [14, 16] give an evidence of our hypothesis about the nature of low-temperature IT peaks in the spectra of YBCO and LSMO ceramics. Like to other cryocrystals, LFIF of solid oxygen was studied using a method of inverse torsion pendulum at frequency 4-8 Hz and oscillation magnitude $\sim 10^{-5}$ through a scheme described earlier [10, 11]. Free samples of solid oxygen were 30 mm in length and 6 mm in diameter. Gas purity was 99.9%, the studies have been conducted at heating and cooling modes.

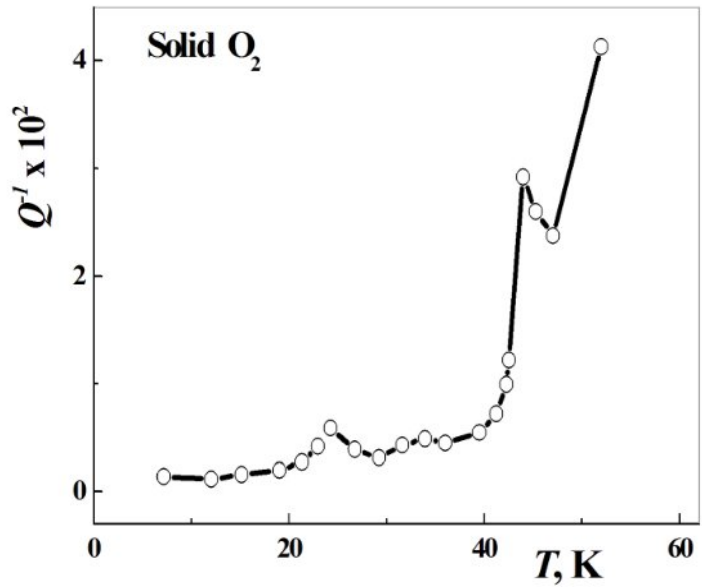

Fig.7 Temperature dependence of LFIF for a sample of crystalline oxygen measured at frequency 4–8 Hz [14].

Fig.7 presents a temperature dependence of LFIF spectra of crystalline oxygen in temperature range 7-52 K. It is seen that IT peaks are observed near the temperatures of phase transitions $T_{\alpha\beta}$ and $T_{\beta\gamma}$. The peaks have considerably different level of relaxation ($Q^{-1}_{\alpha\beta} \sim 10^{-2}$ and $Q^{-1}_{\beta\gamma} \sim 3 \cdot 10^{-2}$). Note that nearly the same relation of the peak magnitudes $Q^{-1}(24\ K)$ and $Q^{-1}(44\ K)$ was observed in the spectrum of HTSC ceramics **$YBa_2Cu_3O_{7-x}$** (see Fig.3). This difference between the peak magnitudes at αβ and βγ phase transitions is attributed to considerable difference in a character of reconstruction of the oxygen crystal lattice under phase transitions. Small hysteresis (3–4 K) was found at heating and cooling which is typical for phase transitions.

Thus, the results of LFIF investigations of free samples of crystalline oxygen give a direct evidence of the hypothesis proposed firstly in [32] about an existence of molecular oxygen in pores of HTSC ceramics and its manifestation in the spectra of such ceramics at low temperatures.

It is worth to note that, in contrast to described above studies of "volume saturation" by oxygen of internal pores in ceramic due to decrease of the oxygen index, a LFIF kinetics of oxide ceramics samples in oxygen atmosphere (i.e. saturation of external open pores) was studied in [15]. It was shown that LFIF kinetics is caused by oxygen absorption on the surface of open pores of the samples which obeys exponent time dependence and reaches saturation after the time about 72 hours. Together with these measurements, a part of closed pores in a studied sample was estimated, which was near 30% from the total porosity. The experiments of [15] demonstrate that the LFIF is a unique tool for the study of active surface state of ceramic samples under absorption or desorption of gases surrounding the studied solids.

## 2. Solid hydrogen in pores of metals

One of the problems of material science is studying of influence of hydrogen saturation on brittle-ductile properties of metals [33]. However, in bulk of the papers this influence was estimated using indirect results that can not give proper information concerning noticeable hydrogen content in pores of the sample.

Direct evidence of hydrogen penetration into internal pores was obtained by Holter et.al [34–38] only. He studied a process of hydrogen saturation of metallic samples through temperature spectra of internal friction (at frequencies 200-300 Hz, i.e. in the LFIF range).

From Fig.8 one can see that below 13 K $\Delta Q^{-1}$ raises in comparison with background value $Q^{-1}$ at T=15 K. At T ≈ 13 K $Q^{-1}$ sharply drops to zero. Naturally, this was explained by melting of solid hydrogen in pores of the metallic samples.

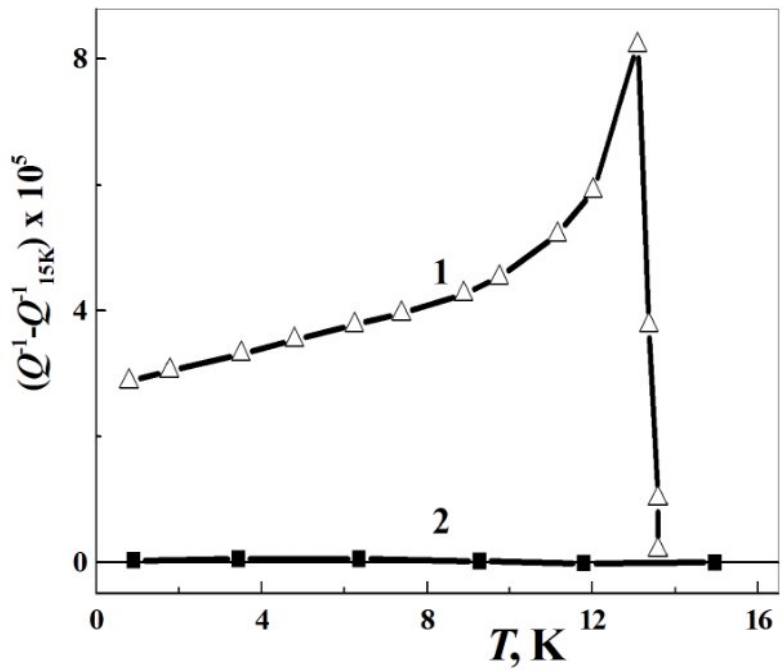

Fig.8. Comparison of temperature dependencies of $Q^{-1}$ for hydrogen saturated sample and the basic (not saturated) one [34–38].
    1 is the curve for saturated sample;
    2 is the curve for not saturated sample.

Note that using the LFIF method the authors of [34–38] not only gave evidence of hydrogen existence in closed pores but thoroughly studied kinetics of hydrogen saturation of the metallic samples (see, for instance, Fig.1–3 in [36].).

## Summary

1. The results of two independent research groups on studying of existence of various gases in pores of solid materials using LFIF method have been presented.
2. We demonstrate that under steeping of the solids in low-temperature media the gases in pores of the solids transform to a crystalline state and cause an additional contribution to the LFIF spectra. Measuring of temperature dependencies of $Q^{-1}$ at this process allows to observe the "cryocrystal" peaks connected with the cryocrystals in pores of the solids.
3. Inasmuch as the peaks temperatures coincide with temperatures of phase transitions of cryocrystals and its triple points then an analysis of the obtained IF spectra allows to identify a composition of gaseous media in pores of the solids.
4. Greatest sensitivity to the contributions of solidified gases in pores of the samples to IF spectra has been observed for the spectra of low-frequency internal friction.
5. So, the LFIF method is a unique structure-sensitive non-destructive studying method to recover and to identify the gases in pores of various solid materials This is of great importance when analyzing an influence of various gaseous media, absorbed in the sample pores, which influence a whole set of physical and thermodynamic properties of a matrix sample.
6. Uniqueness of the LFIF method was noted earlier in [39] and also at analyze of active surface of ceramic HTSC samples during absorption or desorption of gases surrounding the studied samples [15]. Particularly, the LFIF method can be of great importance for studying of cosmic or geologic samples (asteroids, meteorites, rock formations and so on) so it allows to identify a gaseous medium which previously surrounded the studied object.